# CHTW-systems with resource-depended parameters.

# CHTW(R)-systems

**Alexander Yu. Chunikhin**

**Palladin Institute of Biochemistry**

**National Academy of Sciences of Ukraine**

alexchunikhin61@gmail.com

https://orcid.org//0000-0001-8935-0338

**Abstract**. In [1] the concept of CHTW-systems as a multidimensional representation of Petri nets was proposed based on the assumption of multidimensional distribution of tokens (resources) in positions (branes) and, accordingly, multidimensional representation of transitions and arcs. The extension of Petri nets was developed under the assumption of the stationarity of CHTW-system, when its parameters are constant during the system operation. We consider the case when the main parameters of CHTW-system (threshold functions and rate functions) change in accordance with the values of the mark-functions (multidimensional resource) of some container branes of the same CHTW-system. The modification of the basic CHTW-system was designated as a CHTW(R) system, in which (R) means a **R**esource control of the system parameters.

**Keywords**: Petri nets, CHTW-system, resource control, CHTW(R)-system.

1. Introduction

An extension of Petri nets to the case of resources (marking) described by multidimensional functions [1] is developed under the assumption of the stationarity of CHTW-system, when parameters of CHTW-system are constant during the system operation. However, this approach does not allow one to describe the functioning of a wide range of practical systems that are not stationary.

In principle, three ways of control of CHTW-system parameters are possible: internal (autonomous), external and mixed. This paper considers the case of the autonomous resource control of CHTW-system parameters, in which threshold functions and resource uptake intensity (rate) functions change in accordance with the amount of resource in some C-branes to the extent defined by a threshold control matrix and a rate control matrix, respectively.

## 2. Preliminaries

Let's recall the basic concepts, definitions and notations of CHTW-systems [1]. We substitute the basic concepts of Petri nets, which are places, transitions, arrows, with their multidimensional equivalents: branes and carriers.

In this way, C-brane (a **c**ontainer brane) – a resource brane – accumulates, stores and releases distributed *m* resource. T-brane (a **t**ransition brane) — ensures the formation of "firing" (*d*-condition) for both resource uptakes from C-branes with *r* intensity and generation of a new resource. H-carrier (a **th**reshold carrier) – ensures the transfer of the resource from C-brane to T-brane in areas where *m* resource of the C-brane exceeds the value of the carrier *h* threshold. W-carrier (**w**hat/**w**here carrier) – a transformation carrier – ensures the formation of a new resource in C-brane associated with the corresponding T-brane. M – integral (total) resource of CHTW-system – the totality of multidimensional resources of all C-branes in the system.

*Definition 1*. We call *CHTW-system* ($\Xi$) a fivetuple of the following form**:**

$$\Xi = (C, H, T, W, M),$$

where C – a set of C-branes belonging to CHTW-system, T – a set of T-branes belonging to CHTW-system, H – a set of threshold H-carriers from C-branes to T-branes in CHTW-system, W – a set of transformation carriers from T-branes to C-branes in CHTW-system, M – a set of mark-functions that characterize resource distribution in C-branes in CHTW-system.

For the simplest non-stationary CHTW-system (Fig. 1) we write down:

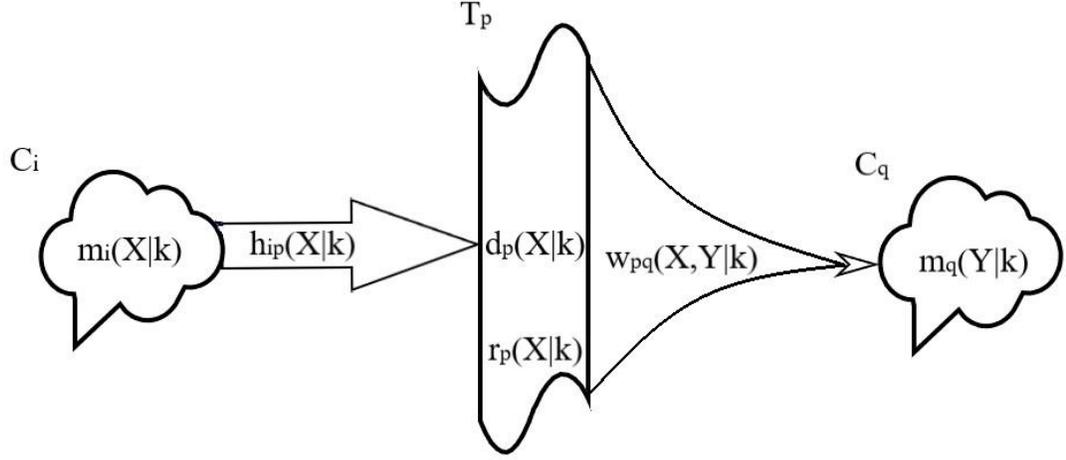

Fig.1

$m_i(X|k)$ – n-dimensional function of $m_i$ resource distribution on $X = (x_1, ..., x_n)$ space in the *i*-th C-brane at the *k*-th step of the CHTW-system operation;

$h_{ip}(X|k)$ – *n*-dimensional function of $h_{ip}$ threshold distribution on $X = (x_1, ..., x_n)$ space in H-carrier between the *i*-th C-brane and the *p*-th T-brane at the *k*-th step of the CHTW-system operation;

$r_p(X|k)$ – *n*-dimensional function of $r_p$ resource uptake intensity distribution from C-branes connected by H-carriers to the *p*-th T-brane on $X = (x_1, ..., x_n)$ space at the *k*-th step of the CHTW-system operation;

$d_p(X|k)$ – the "firing" function of the *p*-th T-brane – *n*-dimensional binary function of "effective" intervals distribution on $X = (x_1, ..., x_n)$ space at the *k*-th step of the CHTW-system operation;

$w_{pq}(X,Y|k)$ – the operator of transformation of the *n*-dimensional "firing" function of the *p*-th T-brane on $X = (x_1, ..., x_n)$ space into the *u*-dimensional function of $m_q$ resource distribution in the *q*-th C-brane on $Y = (y_1, ..., y_u)$ space at the *k*-th step of the CHTW-system operation:

$$w_{pq}(X, Y|k): d_p(X|k) \to m_q(Y|k).$$

The "firing" of the $T_p$-brane is determined by the simultaneous fulfillment of two conditions. At each point $x \in X$, the $m_i(x)$ resource must exceed both the $h_{ip}(x)$ threshold of the corresponding H-carrier and the resource uptake intensity ($r_p(x)$) of the $T_p$-brane. This determines the following:

$\Delta_{ip}(X|k) = m_i(X|k) - h_{ip}(X|k)$ – the function of the resource excess in the $i$-th C-brane over the H-carrier threshold between the $i$-th C-brane and the $p$-th T-brane on $n$-dimensional $X = (x_1, ..., x_n)$ space at the $k$-th step of CHTW-system operation;

$\delta_{ip}(X|k) = m_i(X|k) - r_p(X|k)$ – the function of the resource excess in the $i$-th C-brane over the intensity of the resource uptake by the $p$-th T-brane on $n$-dimensional $X = (x_1, ..., x_n)$ space at the $k$-th step of the CHTW-system operation;

$\Theta(\Delta_{ip}(X|k))$ – the Heaviside function of $\Delta_{ip}(X|k)$ – a distribution of the fact that the resource in the $i$-th C-brane exceeds the H-carrier threshold between the $i$-th C-brane and the $p$-th T-brane on $n$-dimensional $X = (x_1, …, x_n)$ space at the $k$-th step of the CHTW-system operation;

Here the Heaviside function is defined as $\Theta(x) = \begin{cases} 0, x \leq 0 \\ 1, \ x > 0 \end{cases}$;

$\Theta(\delta_{ip}(X|k))$ –the Heaviside function of $\delta_{ip}(X|k)$ – a distribution of the fact that the resource in the $i$-th C-brane exceeds the intensity of the resource uptake by the $p$-th T-brane on $n$-dimensional $X = (x_1, ..., x_n)$ space at the $k$-th step of the CHTW-system operation;

$d_{ip}(X|k) = \Theta(\Delta_{ip}(X|k)) \Theta(\delta_{ip}(X|k))$ – (partial) function of $T_p$-brane "firing" with $C_i$-brane resource.

For the case of multiple entry of two or more C-branes through H-carriers into $T_p$-brane, it can be shown that the integral "firing" function ($d_p(X|k)$) is the product of partial "firing" functions: $d_p(X|k) = \prod(d_{·p}(X|k))$.

For description dynamics of mark-functions we use the following notation, invariant to the special characteristics of branes and carriers: $|m(k)\rangle$ - resource vector-function in C-branes at the $k$-th step of the CHTW-system operation; $\mathbf{H} = \langle h \rangle$ - H-carrier matrix of CHTW-system; $|r\rangle$ - rate vector-function of CHTW-system; $|d(k)\rangle$ - vector-function of T-branes "firing" at the $k$-th step of CHTW-system operation; $\mathbf{W} = \langle w \rangle$ - W-carrier matrix (operator) of CHTW-system. It is also advisable to define *connectivity matrices*: $\mathbf{S}_H = \langle 1_{ct} \rangle$ - (C-branes →T-branes)-connectivity matrix of CHTW-system; $\mathbf{S}_W = \langle 1_{tc} \rangle$ - (T-branes → C-branes)- connectivity matrix of CHTW-system.

Then the $\mathbf{R}_S = \langle r_{ct} \rangle$ *resource uptake matrix* is obtained by

$$R_S = S_H \langle r \rangle = S_H \langle diag(|r\rangle) \rangle.$$

The $\langle diag(|r\rangle)\rangle$ procedure consists of constructing a diagonal matrix $\langle r \rangle$ from the components of the vector $|r\rangle$ in accordance with the rule: $r_i \rightarrow r_{ii}$.

The dynamics of changes in mark-functions (resources) in CHTW-system is described by the following equation:

$$|m(k+1)\rangle = |m(k)\rangle - \mathbf{Rs}\,|d(k)\rangle + \mathbf{W}^T\,|d(k)\rangle.$$

## 3. The concept of CHTW(R)-systems

Let us supplement the basic description of CHTW-system with a control loop for its parameters, in this case, a resource one. Resource control can refer to the regulation of both threshold functions of C-branes and rate functions of T-branes (Fig. 2).

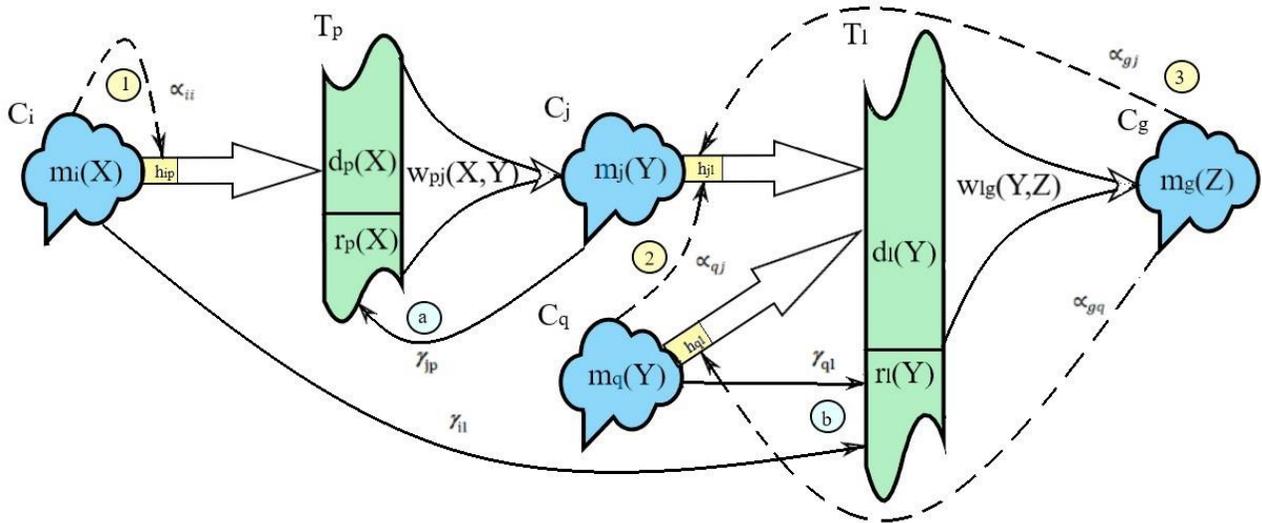

Fig.2

Figure 2 shows possible cases of the control:

(1) – the threshold function of some C-brane is controlled by its own resource function;

(2) – the threshold function of some C-brane is controlled by an "alien" resource function;

(3) – the resource function of a certain C-brane controls the threshold functions of several C-branes;

(a) - the rate function of some T-brane is controlled by the resource function of one C-brane;

(b) - the rate function of some T-brane is controlled by the resource functions of two (several) C-branes.

Case "(2) + (3)" refers to the situation where the threshold function of some C-brane is controlled by the resource functions of several C-branes. Fig. 2 does not show the case where the resource function of some C-brane controls several rate functions of different T-branes.

Let's give a formal definition of CHTW(R)-system as a CHTW-system with controlled parameters.

***Definition 2***. We call *CHTW(R)-system* $\Re$ a seven tuple of the following form:

$$\Re = (C, H, T, W, M, \mathbf{A}, \mathbf{\Gamma}),$$

where C – a set of C-branes belonging to CHTW(R)-system, T – a set of T-branes belonging to CHTW(R)-system, H – a set of threshold H-carriers from C-branes to T-branes in CHTW(R)-system, W – a set of transformation carriers from T-branes to C-branes in CHTW(R)-system, M – a set of mark-functions that characterize resource distribution in C-branes in CHTW(R)-system, $\mathbf{A} = \langle \propto \rangle$ - a threshold control matrix, $\mathbf{\Gamma} = \langle \gamma \rangle$ – a rate control matrix.

The components of the control matrices, for example, some $\alpha_{ij}$ and $\gamma_{gq}$, are operators that associate the mark-function (resource) of $C_i$-brane with the contribution to the change in the threshold function $h_j$ of $C_j$-brane; the mark-function of $C_g$-brane with the contribution to the change in the rate-function $r_q$ of $T_q$-brane, respectively:

$$h_j(k) = h_j(k-1) + \alpha_{ij}(m_i(k)),$$

$$r_q(k) = r_q(k-1) + \gamma_{gq}(m_g(k)).$$

In the simplest (linear) case, when the resource and controllable parameter spaces coincide, we have:

$$h_{j\cdot}(k) = h_{j\cdot}(k-1) + \alpha_{ij} m_i(k),$$

$$r_q(k) = r_q(k-1) + \gamma_{gq} m_g(k).$$

However, such a case rarely occurs in real applications.

The dynamics of changes in mark-functions (resources) in CHTW(R)-system is described by the following equation:

$$|m(k+1)\rangle = |m(k)\rangle + (\boldsymbol{W}^T - \boldsymbol{R}_S(k))|d(k)\rangle.$$

The value of the rate vector at each step of the CHTW(R)-system operation is determined by the sum of its previous value and the value determined by both the mark-function and the rate control matrix

$$|r(k)\rangle = |r(k-1)\rangle + \Gamma^T |m(k)\rangle.$$

The "firing" vector at each step of the CHTW(R)-system operation is defined as

$$|d(k)\rangle = ColProd(\langle d(k)\rangle),$$

where $\langle d(k)\rangle = \langle \Theta(\delta(k))\rangle \bullet \langle \Theta(\Delta(k))\rangle$ is the matrix of partial "firing" functions. The symbol $\bullet$ means the Hadamard product of $\langle \Theta(\delta(k))\rangle$ and $\langle \Theta(\Delta(k))\rangle$ matrices at which

$$d_{ij}(k) = \Theta\left(\delta_{ij}(k)\right) \Theta\left(\Delta_{ij}(k)\right),$$

and $ColProd(\langle d(k)\rangle)$ means the formation of the components of the vector $|d(k)\rangle$ as a product of the elements of the corresponding column.

Suppose $\Theta\langle\delta(k)\rangle = \langle\Theta(\delta(k))\rangle$ which means that the Heaviside function of a matrix is a matrix with the Heaviside function of each element of the matrix.

The matrix function $\langle\delta(k)\rangle$ of the resource excess over the intensity of the resource uptake is defined as

$$\langle\delta(k)\rangle = \boldsymbol{M}_S(k) - \boldsymbol{R}_S(k) = \langle diag(|m(k)\rangle)\rangle\, \boldsymbol{S}_H - \boldsymbol{S}_H\langle diag(|r(k)\rangle)\rangle.$$

The matrix function of the resource excess in C-branes over H-carrier thresholds is defined as

$$\langle\Delta(k)\rangle = \boldsymbol{M}_S(k) - \boldsymbol{H}(k) = \boldsymbol{M}_S(k) - (\boldsymbol{H}(k-1) + \langle diag(\boldsymbol{A}^T|m(k)\rangle)\rangle\, \boldsymbol{S}_H).$$

The $\boldsymbol{R}_S(k)$ resource uptake matrix is obtained by

$$\boldsymbol{R}_S(k) = \boldsymbol{S}_H\langle r(k)\rangle = \boldsymbol{S}_H\langle diag(|r(k)\rangle)\rangle.$$

The final expression for the dynamics of changes in mark-functions (resources) in CHTW(R)-system takes the form

$$|m(k+1)\rangle = |m(k)\rangle + (W^T - S_H\langle diag(|r(k)\rangle)\rangle) ColProd(\langle \Theta(\delta(k))\rangle \bullet \langle \Theta(\Delta(k))\rangle).$$

## 4. An example of CHTW(R)-system description

Let's find mathematical expressions that describe the CHTW(R)-system operation at the k-th step using the example of the CHTW(R)-system shown in Fig. 2.

The vectors of resource distribution and resource uptake (rates) are respectively

$$|m(k)\rangle = \begin{pmatrix} m_i(k) \\ m_j(k) \\ m_q(k) \\ m_g(k) \end{pmatrix}, \quad |r(k)\rangle = \begin{pmatrix} r_p(k) \\ r_l(k) \end{pmatrix}.$$

The C-branes → T-branes connectivity matrix ($S_H$) and the T-branes → C-branes connectivity matrix ($S_W$) of the CHTW(R)-system are respectively

$$S_H = \begin{pmatrix} 1 & 0 \\ 0 & 1 \\ 0 & 1 \\ 0 & 0 \end{pmatrix}, \quad S_W = \begin{pmatrix} 0 & 1 & 0 & 0 \\ 0 & 0 & 0 & 1 \end{pmatrix}.$$

The H-carrier matrix and the W-carrier matrix of the CHTW(R)-system are respectively

$$H(k) = \begin{pmatrix} h_{ip}(k) & 0 \\ 0 & h_{jl}(k) \\ 0 & h_{ql}(k) \\ 0 & 0 \end{pmatrix}, \quad W = \begin{pmatrix} 0 & w_{pj} & 0 & 0 \\ 0 & 0 & 0 & w_{lg} \end{pmatrix}.$$

The threshold control matrix $A$, and the rate control matrix $\Gamma$ have the following forms

$$A = \begin{pmatrix} \alpha_{ii} & 0 & 0 & 0 \\ 0 & 0 & 0 & 0 \\ 0 & \alpha_{qj} & 0 & 0 \\ 0 & \alpha_{gj} & \alpha_{gq} & 0 \end{pmatrix}, \quad \Gamma = \begin{pmatrix} 0 & \gamma_{il} \\ \gamma_{jp} & 0 \\ 0 & \gamma_{ql} \\ 0 & 0 \end{pmatrix}.$$

Then, the rate-vector of the CHTW(R)-system takes the form

$$|r(k)\rangle = |r(k-1)\rangle + \Gamma^T |m(k)\rangle = |r(k-1)\rangle + \begin{pmatrix} 0 & \gamma_{jp} & 0 & 0 \\ \gamma_{il} & 0 & \gamma_{ql} & 0 \end{pmatrix} \begin{pmatrix} m_i(k) \\ m_j(k) \\ m_q(k) \\ m_g(k) \end{pmatrix}$$

$$= \begin{pmatrix} r_p(k-1) + \gamma_{jp}(m_j(k)) \\ r_l(k-1) + \gamma_{il}(m_i(k)) + \gamma_{ql}(m_q(k)) \end{pmatrix}.$$

The resource uptake matrix $R_S(k)$ is

$$R_S(k) = \begin{pmatrix} r_p(k-1) + \gamma_{jp}(m_j(k)) & 0 \\ 0 & r_l(k-1) + \gamma_{il}(m_i(k)) + \gamma_{ql}(m_q(k)) \\ 0 & r_l(k-1) + \gamma_{il}(m_i(k)) + \gamma_{ql}(m_q(k)) \\ 0 & 0 \end{pmatrix}.$$

The matrix of the resource excess over the intensity of the resource uptake takes the form

$$\langle \delta(k) \rangle = \begin{pmatrix} m_i(k) - r_p(k-1) - \gamma_{jp}(m_j(k)) & 0 \\ 0 & m_j(k) - r_l(k-1) - \gamma_{il}(m_i(k)) - \gamma_{ql}(m_q(k)) \\ 0 & m_q(k) - r_l(k-1) - \gamma_{il}(m_i(k)) - \gamma_{ql}(m_q(k)) \\ 0 & 0 \end{pmatrix}.$$

The matrix of the resource excess over the threshold takes the form

$$\langle \Delta(k) \rangle$$

$$= \begin{pmatrix} m_i(k) & 0 \\ 0 & m_j(k) \\ 0 & m_q(k) \\ 0 & 0 \end{pmatrix} - \begin{pmatrix} h_{ip}(k-1) & 0 \\ 0 & h_{jl}(k-1) \\ 0 & h_{ql}(k-1) \\ 0 & 0 \end{pmatrix} - \langle diag \begin{pmatrix} \alpha_{ii} & 0 & 0 & 0 \\ 0 & 0 & \alpha_{qj} & \alpha_{gj} \\ 0 & 0 & 0 & \alpha_{gq} \\ 0 & 0 & 0 & 0 \end{pmatrix} \begin{pmatrix} m_i(k) \\ m_j(k) \\ m_q(k) \\ m_g(k) \end{pmatrix} \rangle S_H$$

$$= \begin{pmatrix} m_i(k) - h_{ip}(k-1) - \alpha_{ii}(m_i(k)) & 0 \\ 0 & m_j(k) - h_{jl}(k-1) - \alpha_{qj}(m_q(k)) - \alpha_{gj}(m_g(k)) \\ 0 & m_g(k) - h_{ql}(k-1) - \alpha_{gq}(m_g(k)) \\ 0 & 0 \end{pmatrix}.$$

Then we obtain the final expression for the dynamics of mark-functions changes in the CHTW(R)-system (Fig.2):

$$|m(k+1)\rangle = |m(k)\rangle + \begin{pmatrix} -r_p(k-1) - \gamma_{jp}(m_j(k)) & 0 \\ w_{pj} & -r_l(k-1) - \gamma_{il}(m_i(k)) - \gamma_{ql}(m_q(k)) \\ 0 & -r_l(k-1) - \gamma_{il}(m_i(k)) - \gamma_{ql}(m_q(k)) \\ 0 & w_{lg} \end{pmatrix}$$

$$\cdot ColProd(\langle \Theta(\delta(k)) \rangle * \langle \Theta(\Delta(k)) \rangle).$$

## 5. Conclusion

These intermediate and final expressions for the dynamics of mark-functions in CHTW(R)-system are obtained under the assumption that the spaces of all branes and carriers are consistent. Since the elements of the control matrices are operators, this gives grounds to assert that some parameter over the space X can be controlled by a resource over the space Y, Y≠X. This state of affairs is quite adequate for most applications, which makes CHTW(R)-system an effective tool for simulating real multidimensional systems. In the linear case, the spaces of the resource and the parameter controlled by it must coincide.

It is advisable to carry out further research in the field of CHTW(R)-systems in the directions of both the dynamics of CHTW(R)-systems under external control and the adaptive behavior of such systems.